\documentclass[aps,pre,groupedaddress,twocolumn]{revtex4}
\usepackage{CJKutf8}

\usepackage[utf8]{inputenc}
\usepackage{bm,color,amsmath,amssymb,siunitx,blindtext}
\usepackage{graphicx,epstopdf}
\usepackage{lipsum}

\newcommand{\D}{\mathrm{d}}
\newcommand{\E}{\mathrm{e}}
\newcommand{\PAM}{\mathrm{PAM}}
\newcommand{\bp}{\mathrm{bp}}

\begin{document}
\begin{CJK*}{UTF8}{gbsn}


\title{Search and localization dynamics of the CRISPR/Cas9 system}


\author{Qiao Lu (路桥)$^1$}
\author{Deepak Bhat$^1$}
\author{Darya Stepanenko$^{1,2,3}$} 
\author{Simone Pigolotti$^1$}
\email{simone.pigolotti@oist.jp}
\affiliation{$^1$Biological Complexity Unit, Okinawa Institute of Science and Technology Graduate University, Onna, Okinawa 904-0495, Japan,}
\affiliation{$^2$Laufer Center for Physical and Quantitative Biology, Stony Brook University, New York, USA,}
\affiliation{$^3$Department of Applied Mathematics and Statistics, Stony Brook University, New York, USA.}
\date{\today}

\begin{abstract}
The CRISPR/Cas9 system acts as the prokaryotic immune system and has important applications in gene editing. The protein Cas9 is one of its crucial components. The role of Cas9 is to search for specific target sequences on the DNA and cleave them.  In this Letter, we introduce a model of facilitated diffusion for Cas9 and fit its parameters to single-molecule experiments. Our model confirms that Cas9 search for targets by sliding, but shows that its sliding length is rather short. We then investigate how Cas9 explores a long stretch of DNA containing randomly placed targets. We solve this problem by mapping it into the theory of Anderson localization in condensed matter physics. Our theoretical approach rationalizes experimental evidences on the distribution of Cas9 molecules along the DNA. 
\end{abstract}


\maketitle
\end{CJK*}

The discovery of the CRISPR/Cas9 system has revealed the functioning of the bacterial immune response and has opened previously unimaginable possibilities for gene editing \cite{jinek2012programmable}.  The protein Cas9 is a central actor in this system. Cas9 is an endonuclease that is able to load a guide RNA strand. Its target is a sequence on the DNA complementary to the guide RNA, which Cas9 can identify and cleave. This recognition process has attracted considerable attention, including from the modeling side \cite{farasat2016biophysical,shvets2017mechanism,klein2018hybridization}. Recognition is triggered by a three-base sequence called protospacer adjacent motif (PAM), which precedes the target. The first two bases of a PAM  are guanine, whereas the third can be any base. Cas9 can transiently bind to a PAM even in the absence of a neighboring target \cite{jones2017kinetics,globyte2019crispr}. 

Other proteins such as the Lac repressor in Escherichia coli  \cite{hammar2012lacFD}  find their targets along the DNA by a mechanism termed facilitated diffusion -- an alternance of 3D diffusion in the cytosol and one-dimensional diffusive sliding along the DNA chain \cite{berg1981diffusion}. This mechanism can significantly improve search efficiency \cite{berg1981diffusion,mirny2009protein}.  The theory of facilitated diffusion has been extended to take into account the energetics of target search along the DNA \cite{slutsky2004kinetics,bauer2015real,cencini2018energetic} and other processes such as hopping, i.e., the possibility for proteins to briefly detach from DNA and then reattach at short distance \cite{lomholt2009facilitated}. This notion has stimulated experimental efforts to determine whether Cas9 finds its target by facilitated diffusion as well \cite{sternberg2014dna,singh2016real,globyte2019crispr}.

However, experimental single-molecule studies using DNA curtains \cite{sternberg2014dna} and fluorescence resonance energy transfer (FRET) \cite{singh2016real} did not find evidences of sliding. They however found that the lifetime of Cas9 binding events is well fitted by a double exponential even in the absence of targets, suggesting a complex binding mechanism.  In contrast, a more recent FRET experimental study provides evidences that Cas9 can slide \cite{globyte2019crispr}.  This study found that, in a DNA sequence containing multiple PAMs without targets, the two exponential constants characterizing the binding lifetime distribution depend on the number of PAMs and the distance between them.  In the absence of PAMs, this distribution reduces to a single exponential.  The dependence of the exponential constants on the distance between PAMs suggests that the characteristic sliding length of Cas9 falls below the spatial resolution of previous experiments \cite{sternberg2014dna}, potentially explaining why sliding was not previously observed.

An alternative way of probing the search dynamics of Cas9 is to experimentally  measure the distribution of Cas9 molecules bound along the DNA. For example, an experiment based on DNA curtains shows that Cas9 is localized in regions that extend for hundreds of base pairs length around targets \cite{sternberg2014dna}. This length scale is much larger than the sliding length suggested by Ref. \cite{globyte2019crispr}.

These contrasting experimental evidences call for a theoretical explanation. To this aim, it is useful to think about a bacterial genome as a long stretch of DNA in which a large number of PAMs are disorderly distributed. For comparison, the E.coli genome is 4.6 million base pairs long and contains about half million PAMs \cite{jones2017kinetics}. We want to estimate the typical localization length of Cas9 on such DNA sequences. This problem bears an analogy with the theory of Anderson localization \cite{anderson1958absence}. This theory predicts that, under general conditions, eigenvectors of disordered one-dimensional diffusive systems are localized, with profound consequences for fields of physics ranging from condensed matter to disordered and chaotic systems \cite{book}. In this analogy, PAMs play the role of defects in one dimensional lattices.

In this Letter, we show that a facilitated diffusion model quantitatively explains the dynamics of Cas9 observed in single-molecule experiments. We then formalize the mapping between facilitated diffusion and Anderson localization. This approach permits us to determine the localization length of Cas9 on typical long DNA strands and explain the discrepancy between the sliding length in \cite{globyte2019crispr} and the localization length in \cite{sternberg2014dna} in terms of a hopping mechanism. The mapping presented in this Letter can be used to study the dynamics of other DNA binding proteins, such as transcription factors.

\begin{figure}[htb!]
\centering
\includegraphics[width=0.9\linewidth]{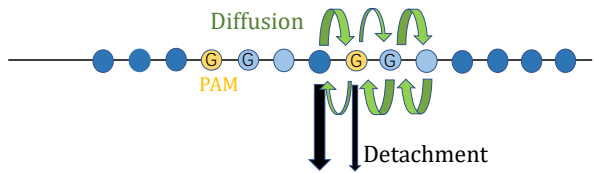}
\caption{Scheme of the model. PAM sites and non-specific sites are shown in yellow and blue, respectively. The second and third bases of PAM sequences are considered as non-specific sites (light blue). Green arrows represent sliding rates and black arrows represent unbinding rates, see Eq.~\eqref{eq:rates}. Thicker arrows correspond to larger rates.  }
\label{fig:1}
\end{figure}

We consider a Cas9 protein that binds on a DNA chain of length $N$ and slides along it before detaching, see Fig.~\ref{fig:1}.  Our aim is to quantify the distribution of duration of binding events depending on the arrangement of specific PAM sites along the DNA chain. We introduce the probability $p_n(t)$ that Cas9 is bound at site $n$ at time $t$, given that it had attached on the DNA at time $t=0$. Each site represents a nucleotide position $n=1\dots N$. We assume attachment to be non specific, so that $p_n(t=0)=1/N$. 

We distinguish between two types of DNA sites. PAM sites are those at the beginning of a PAM sequence, where Cas9 can bind specifically. We consider every other site as non-specific, including the two other base pairs constituting a PAM, see Fig.~\ref{fig:1}. We call $E_n$ the binding energy of Cas9 at position $n$. We assume that all non-specific sites have the same binding energy $E_n=0$. If $n$ is a PAM site, then $E_n=-\beta$, with $\beta>0$. All energies are expressed in units of $k_\mathrm{B}T$, where $k_\mathrm{B}$ is the Boltzmann constant and $T$ the temperature. Our aim is to analyze single binding events and therefore we do not consider rebinding after detachment.

The probabilities $p_n(t)$ evolve according to the master equation
\begin{align}
  \frac{\D}{\D t}p_{n}(t)=&D_{n,n+1}p_{n+1}+D_{n,n-1}p_{n-1}\nonumber\\
  &-(D_{n+1,n}+D_{n-1,n}+k_n)p_{n} ,
  \label{equation:-1}
\end{align}
in which $D_{n,m}=D e^{E_m}$ and $k_n=k e^{E_n}$, where the diffusion rate $D$ and the unbinding rate $k$ are given parameters. We impose vanishing fluxes at the boundaries, $D_{0,1}=D_{1,0}=D_{N,N+1}=D_{N+1,N}=0$. This choice of rates satisfies the detailed balance condition $D_{n,m} e^{-E_m}=D_{m,n}e^{-E_n}$.

We express the model in vector notation by defining $\mathbf{p}(t)=(p_1(t),p_2(t)\dots p_N(t))$. We write Eq.~\eqref{equation:-1} as $\D \mathbf{p}/ \D t = \hat{A}\mathbf{p}$, where the elements $A_{m,n}$ of the matrix $\hat{A}$ are given by
\begin{equation}\label{eq:rates}
A_{m,n}=\left\{
\begin{array}{lr}
De^{E_n} & \mbox{if}\ |n-m|=1\\
-(k+2D)e^{E_m} & \mbox{if}\ n=m .
\end{array}
\right.
\end{equation}
The formal solution to the master equation is $\mathbf{p}(t)=e^{\hat{A}t}\mathbf{p}(0)$, where $\mathbf{p}(0)$ is the uniform initial condition. The eigenvalue equation associated with the master equation is
\begin{equation}\label{eq:eigenvalues}
\hat{A}\boldsymbol{\psi} = -\lambda \boldsymbol{\psi}.
\end{equation}
Equation~\eqref{eq:eigenvalues} is solved by a set of eigenvalues $\lambda=\lambda_1,\lambda_2,\dots \lambda_N$ and associated right eigenvectors $\boldsymbol{\psi}=\boldsymbol\psi^{(1)},\boldsymbol\psi^{(2)},\dots \boldsymbol\psi^{(N)}$, assumed to be normalized. The solution of the master equation can be decomposed into eigenvalues
\begin{equation}\label{eq:eigen_expansion}
    \mathbf{p}(t)=\sum_{i=1}^N e^{-\lambda_i t} c_i \boldsymbol{\psi}^{(i)} ,
\end{equation}
where the coefficients $c_i$ are determined by the initial condition. Because of detachment, one has $\lim_{t\rightarrow\infty}p_i(t)=0$ for all $i$. This fact and the detailed balance condition imply that all eigenvalues must be real and positive. We sort the eigenvalues so that $\lambda_1$ is the smallest one.

The total probability that Cas9 is still bound at a time $t$ is given by $P(t)=\sum_n p_n(t)$.  Since we are considering a single binding event, $P(t)$ is a decreasing function of $t$. We define the instantaneous detachment rate $g(t)=-\D/\D t\, P(t)$. Single-molecule experiments \cite{sternberg2014dna,singh2016real,globyte2019crispr} observed that the temporal decay of $g(t)$, and therefore of $P(t)$, is characterized by two distinct exponential slopes at short and long times. 

To understand these two regimes, we  focus on $P(t)$ and define its instantaneous exponential slope $K(t)=-\D/\D t\, \ln P(t)$. We also define  the total probability $P_{\PAM}(t)=[\sum_{n\in \PAM}p_n(t)]/P(t)$ of Cas9 being bound to a PAM site at time $t$, given that it had not detached yet. By summing Eq.~\eqref{equation:-1} over $n$, we find that 
\begin{equation}\label{eq:slope}
   K(t)= k\left[1-P_\PAM(t)\right] +k \E^{-\beta} P_{\PAM}(t) .
\end{equation}
Considering that $0\le P_{\PAM}(t)\le 1$,  the slope $K(t)$ is limited by the two unbinding rates:
\begin{equation}\label{eq:sloperange}
k\E^{-\beta} \le K(t)  \le k.
\end{equation}

The value of $K(t)$ in this range is determined by $P_\PAM(t)$. Since the initial distribution is uniform, at short times $P_{\PAM}$ is equal to the fraction of PAM sites. Given that this fraction is usually small, Eq.~\eqref{eq:slope} implies $K(t)\approx k$ at short times.  In the long time limit,  Eq.~\eqref{eq:eigen_expansion} leads to conclude that  $K(t)=\lambda_1$. 

\begin{figure}[htb]
\includegraphics[width=1.0\linewidth]{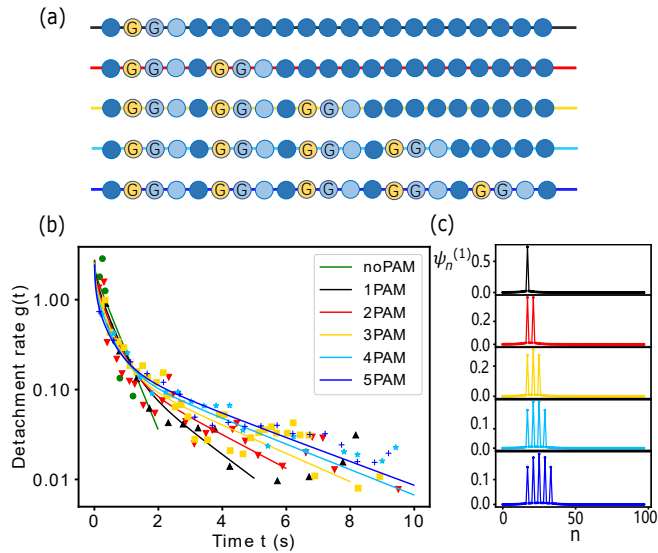}
\caption{(a) Arrangements of PAM sites used in the experiments in \cite{globyte2019crispr}. Line colors correspond to the different curves in panel b. The figure shows only the portion of the DNA sequence of length $N=98$ where the PAM sites are located. (b) Comparison of the prediction of our model (lines) with experiments \cite{globyte2019crispr} (points). Model parameters are determined by jointly fitting the experimental data for $j=0\dots 5$ PAM sites using maximum likelihood, see Appendix A. (c) Eigenvectors $\boldsymbol{\psi}^{(1)}$ for $j=1\dots 5$. } 
\label{fig:2}
\end{figure}

Experiments in \cite{globyte2019crispr} measured the distribution of Cas9 binding events on DNA sequences containing from $0$ to $5$ PAM sites . We jointly fitted the parameters $k$, $\beta$, and $D$ to these six experiments, see Fig.~\ref{fig:2}a.  Solutions of the master equation~\eqref{equation:-1} with the best-fit parameters reproduce the double exponential behavior and fit well the experimental data, see Fig.~\ref{fig:2}b. The fitted values of the parameters are $k=1.94\pm 0.10\, s^{-1}$, $\beta=3.34\pm 0.07$, and $D=52\pm 9\, \bp^2 s^{-1}$. Experiments on a different variant of Cas9 find differences in binding energy between PAM and near-cognate sites that are comparable with our estimate of $\beta$ \cite{farasat2016biophysical}. A more detailed model where each non-PAM sequence is characterized by a different binding energy, leads to similar fitted values of the corresponding rates, see Appendix B. These evidences support robustness of our results.

At increasing number of PAM sites, the second slope in Fig.~\ref{fig:2}a becomes significantly less steep than the first. According to Eq.~\eqref{eq:slope}, this means that, at long times, Cas9 is much more localized on PAM sites compared with short times.  Inspecting the eigenvectors $\boldsymbol{\psi}^{(1)}$ associated with the smallest eigenvalue $\lambda_1$ confirms this idea, see Fig.~\ref{fig:2}c.

To gain further insight into the dynamics observed in Fig.~\ref{fig:2}b,  we analytically compute $\lambda_1$ and its eigenvector for an infinitely long chain with a single PAM site at $n=0$.  For $|n|>1$, the eigenvector satisfies
\begin{equation}\label{eq:me_onepam}
-\lambda_1 \psi^{(1)}_n = D\left(\psi^{(1)}_{n+1}+\psi^{(1)}_{n-1}-2\psi^{(1)}_{n}\right)-k\psi^{(1)}_{n} .
\end{equation}

We assume a solution of the form $\psi_{n}\propto e^{-|n|/\ell}$ where $|n|>0$ and we define $\ell$ as the sliding length. Substituting into Eq.~\eqref{eq:me_onepam} we obtain
\begin{equation}\label{eq:dispersion}
  k-\lambda_1 = 2D[\cosh (1/\ell)-1].
\end{equation}
By expanding the $\cosh$ at first order we find $\ell\approx\sqrt{D/(k-\lambda_1)}$. Note that $\lambda_1\le k$ due to Eq.~\eqref{eq:sloperange}. The three unknown $\lambda_1$, $\ell$ and $\psi_0$ can be determined from Eq.~\eqref{eq:dispersion} and the equivalents of Eq.~\eqref{eq:me_onepam} for $n=0$ and $|n|=1$. Substituting the fitted parameters of Fig.~\ref{fig:2}, we find $\ell\approx 6.2$ bp. 

\begin{figure}[htb]
\includegraphics[width=0.8\linewidth]{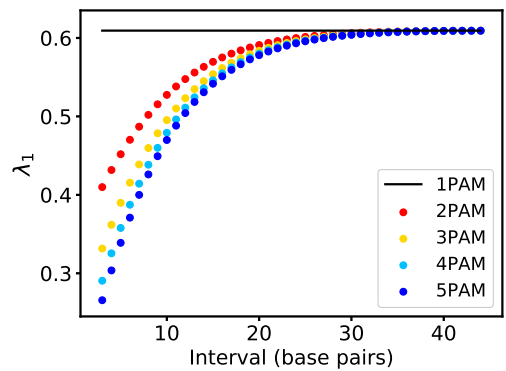}
\caption{Interference between $j=1\dots 5$ equally spaced PAM sites on an infinite DNA chain. Lowest eigenvalue $\lambda_{1}$ as a function of the interval between the PAM sites. Points are obtained by numerically diagonalizing the matrix $\hat{A}$ corresponding to each case, with $N=220$. The horizontal line marks the value of $\lambda_{1}$ for a single PAM sequence, from the solution of Eq.~\eqref{eq:me_onepam}.}
\label{fig:3}
\end{figure}

Both our model and experiments \cite{globyte2019crispr} show that the lifetime of long binding events increases at increasing number of PAMs, see Fig.~\ref{fig:2}b. In the model, this means that $\lambda_1$ is a decreasing function of the number of PAMs.  This effect can be explained by interference among PAM sites, i.e. the fact that the eigenvector $\boldsymbol{\psi}^{(1)}$ for $j$ PAM sites is not simply a superimposition of $j$ single-PAM eigenvectors, unless the interval between the PAM sites is much larger than $\ell$. Only in this limit  binding events around each PAM site behave independently, and the long-time exponential slope becomes independent of the number of PAM sites, see Fig.~\ref{fig:3}. At shorter intervals, interference leads to an increase in target occupancy. This implies that, at large $t$, $P_\PAM(t)$, and therefore the typical lifetime of binding events $1/\lambda_1$, are decreasing functions of the interval between the PAM sites, see Eq.~\eqref{eq:slope} and Fig.~\ref{fig:3}.

In summary, we found that the distribution of a Cas9 molecule in a region of DNA containing several PAM sites tends to be localized. We now study the behavior of Cas9 on a very long stretch of DNA including a disordered assortment of PAM sites.  The theory of Anderson localization predicts that, in such disordered one-dimensional systems, eigenvectors are exponentially localized:
\begin{equation}\label{eq:def_localizationlength}
\psi_n \sim e^{-\frac{|n-n^*|}{\gamma(\lambda)}},
\end{equation}
where $n^*$ is the location of the eigenvector peak and $\gamma(\lambda)$ is the localization length associated with the eigenvalue $\lambda$. The localization length $\gamma$ can be thought as the generalization of the sliding length $\ell$: the former is defined for an arbitrary disordered DNA chain, whereas the latter is defined for a single target. Our hypothesis is that the localization length associated with the smallest eigenvalues of Cas9 dynamics can explain the results of DNA curtains experiments \cite{sternberg2014dna}.

We sharpen the analogy between our problem and the Anderson localization by rescaling the components of our eigenvectors by the Boltzmann weight, $f_{n}=\psi_{n} \exp(E_{n})$. With this transformation, Eq.~\eqref{eq:eigenvalues} assumes the same form for PAM and non-PAM sites:
\begin{equation}
  f_{n+1}+f_{n-1}-\left(2+\frac{k-\lambda e^{-E_{n}}}{D}\right)f_{n}=0.
  \label{equation:master}
\end{equation}
This equation is formally similar to the discrete Schr\"{o}dinger equation in Anderson's original work \cite{anderson1958absence}. It can be solved by the transfer matrix method. We introduce the vector $\mathbf{f}_n=(f_{n},f_{n-1})$ and the transfer matrix
\begin{equation}
\hat{T}_n=
\left( \begin{array}{cc}
2+\frac{k-\lambda e^{-E_{n}}}{D} & -1 \\
1 & 0
\end{array} \right).
\label{equation:transfer}
\end{equation}
With these definitions, we rewrite Eq.~\eqref{equation:master} as
\begin{equation}
    \mathbf{f}_{n+1}=\hat{T}_{n} \mathbf{f}_{n}
\end{equation}
and therefore
\begin{equation}\label{eq:prodrand}
    \mathbf{f}_{N}=\prod_{n=1}^{N-1}\hat{T}_{n}\mathbf{f}_{1}.
\end{equation}
We assume that, in a typical long DNA sequence, each site $n$ has a probability $1/16$ to be a PAM site, thereby affecting the value of $E_n$ in the corresponding matrix $T_n$.  In this view, Eq.~\eqref{eq:prodrand} expresses the solution of the eigenvalue equation as a product of random matrices \cite{book}.

The localization length $\gamma$ can be calculated from this product with an approach proposed by Herbert, Jones, and Thouless \cite{herbert1971localized,thouless1972relation}. This approach rests on the idea that $f_N(\lambda)$, with appropriate boundary conditions, vanishes if $\lambda$ is an eigenvalue and changes sign as a function of $\lambda$ at every eigenvalue. This argument leads to the expression
\begin{equation}
  \frac{1}{N}\ln f_{N}(\lambda)=\frac{1}{N}\sum_{n=1}^{N-1}\ln|\lambda_{n}-\lambda|+\frac{i\pi}{N}\sum_{n=1}^{N-1}\theta(\lambda-\lambda_{n})+\frac{1}{N}\ln A
  \label{equation:thoul1}
\end{equation}
where $\theta$ is the Heaviside step function and $A$ is a finite constant. Taking the limit $N\rightarrow \infty$, we define
\begin{equation}
 \Lambda(\lambda)= \lim_{N\rightarrow \infty}\frac{1}{N}\ln f_{N}(\lambda)=\lim_{N\rightarrow \infty}\frac{1}{N}\ln\left(\mathrm{Tr}\prod_{n=1}^{N}\hat{T}_{n}\right),
  \label{equation:thoul2}
\end{equation}
The Furstenberg theorem guarantees that $\Lambda(\lambda)$ is independent of the realization of the disorder and of the choice of $\bf{f}_1$ \cite{ishii1973localization, furstenberg}. 

\begin{figure}[hbt!]
\centering
\includegraphics[width=1.0\linewidth]{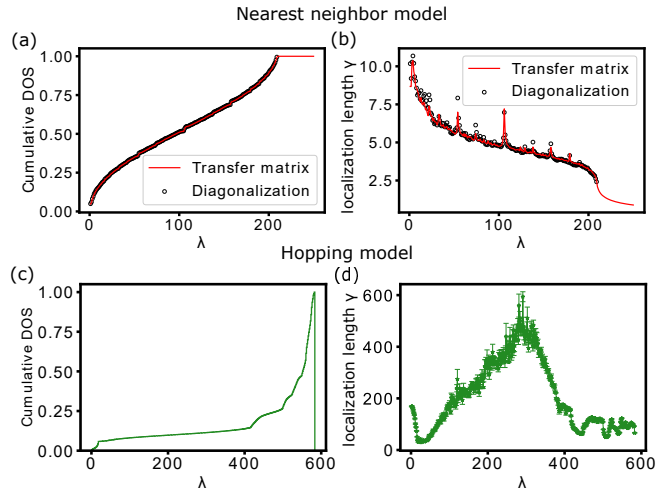}
\caption{(a) Cumulative density of states (DOS) and (b) localization length as function of $\lambda$ for the nearest neighbour model,  Eq.~\eqref{equation:-1}, computed using Eq.~\eqref{equation:thoul1}.  Results obtained by the transfer matrix method agree with those obtained by direct diagonalization. The DNA chain length is $N=10^6$ for the transfer matrix method and $N=5000$ for the direct diagonalization. (c) Cumulative DOS and (d) localization length for the hopping model expressed by Eq.~\eqref{eq:hoppingmodel}, computed using Eq.~\eqref{eq:directlocal}.  In this case, the DNA chain length is $N=2000$.}
\label{fig:4}
\end{figure}

The inverse of the real part of $\Lambda(\lambda)$ can be identified with the localization length $\gamma(\lambda)$ thanks to a result known as the Borland conjecture \cite{borland}. The validity of this conjecture for our class of systems is supported by numerical and theoretical studies \cite{ishii1973localization, matsuda1970localization}.  Further, Eq.~\eqref{equation:thoul1} links the imaginary part of $\Lambda$ with the cumulative density of states. Computing $\Lambda(\lambda)$ from the product of transfer matrices, we find that the localization length for the whole spectrum is always shorter than 11 base pairs,  see Fig.~\ref{fig:4}b. 

We remark that the disordered arrangement of PAM sites is crucial for this result. In a long DNA chain containing a periodic arrangement of PAM sites, the eigenvectors are extended rather than localized, see Appendix C. 

The localization lengths in Fig.~\ref{fig:4} are much shorter than those observed in DNA curtains experiments \cite{sternberg2014dna}. We assume that this discrepancy can be explained by the following idea. Measuring the distribution of Cas9 in an experiment amounts to performing an ``ensemble average'' which is potentially affected by search mechanisms other than sliding (such as hopping). In contrast, FRET experiments focus on individual sliding events, which are unaffected by such mechanisms. 

To test this idea, we generalize our model to include hopping. In a hopping event, Cas9 detaches and then reattaches to the DNA at a short distance. This amounts to include in our master equation diffusion among non-nearest neighboring sites:
\begin{equation}\label{eq:hoppingmodel}
D_{m,n}=De^{E_n}h(|n-m|) ,
\end{equation}
where $h(n)$ is a positive decreasing function characterizing the probability of hopping events at a given distance $n$ relative to sliding events. We impose $h(1)=1$, so that nearest-neighbor sliding is consistent with Eq.~\eqref{eq:rates}. We determine the function $h(n)$ from the solution of a diffusion equation in cylindrical coordinates, see \cite{lomholt2009facilitated} and Appendix D. Unbinding rates in the hopping model are the same as in Eq.~\eqref{eq:rates}. For models with next to nearest neighbor interactions, such as our hopping model, the localization length can not be computed using Eqs.~\eqref{equation:thoul1} and \eqref{equation:thoul2}, see \cite{biddle2011localization}. We therefore estimate the localization length by a more direct strategy, although computationally heavier. Assuming that a given eigenvector $\boldsymbol{\psi}^{(i)}$ associated with an eigenvalue $\lambda_i$ is localized, we obtain from Eq.~\eqref{eq:def_localizationlength} that
\begin{equation}
    \gamma (\lambda_i)\sim-\frac{(N-1)}{\ln\left[\psi_1^{(i)}\psi_N^{(i)}\right]}.
    \label{eq:directlocal}
\end{equation}
In this case, the localization length associated with the lowest eigenvalues is on the same order of the experimentally measured one (hundreds of base pairs, see Fig.~\ref{fig:4}d).

In conclusion, in this Letter we studied the search dynamics of Cas9 along the DNA. We have shown that the predictions of a facilitated diffusion model with a short sliding length are consistent with the result of single-molecule FRET experiments. By applying the theory of Anderson localization, we have argued that a hopping mechanism can explain how Cas9 is generically distributed along the DNA. 

The mapping to  Anderson localization introduced in this Letter is a powerful tool that can be applied to any protein performing facilitated diffusion, such as transcription factors. Modern immunoprecipitation techniques permit to measure binding profiles of transcription factors along the DNA at the base pair resolution \cite{rhee2011comprehensive}. However, the interpretation of these binding profiles is still under debate \cite{macquarrie2011genome}. Our approach can be combined with sequence-dependent models of facilitated diffusion by transcription factors \cite{slutsky2004kinetics,bauer2015real,cencini2018energetic} to shed light on this crucial problem in biophysics.

\begin{acknowledgments}
We are grateful to Chirlmin Joo and Viktorija Globyte for sharing experimental data.
\end{acknowledgments}

\appendix

\section{Maximum likelihood fit }\label{sec:likelihood}

\begin{figure}[htb!]
\centering
\vspace*{0.1em}
\includegraphics[width=1.0\linewidth]{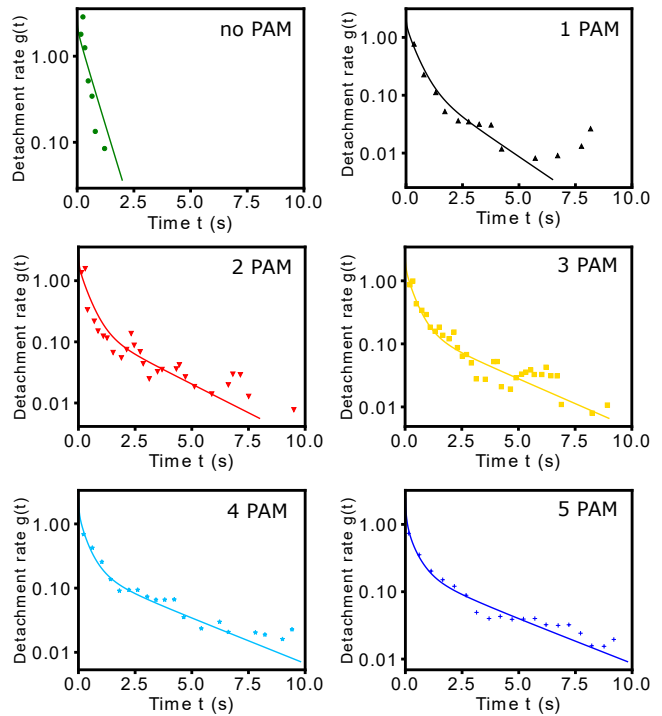}
\caption{Fitted detachment rate as a function of time for different number of PAMs. Curves and data are the same as in Fig.~2b of the Main Text, but presented in separate panels.}
\label{fig:s1}
\end{figure}

To fit the experimental data from \cite{globyte2019crispr}, we express the likelihood of one specific experiment as
\begin{equation}\label{eq:mle}
   \mathcal{L}_j = \mathcal{N}_j!\prod_{i}\frac{\rho_{i,j}^{n_{i,j}}}{n_{i,j}!}
\end{equation}
where the index $0\le j\le5$ indicates the number of PAM sites in each experiment. For each experiment $j$, we call $n_{i,j}$  the number of binding events in the $i$th bin of the histogram,  $\mathcal{N}_j=\sum_i n_{i,j}$ is the total number of binding events,  and $\rho_{i,j}=P(t_{i-1})-P(t_i)$ is the probability that the duration of a binding event falls into the $i$th bin. This probability is obtained from numerical integration of Eq.~(1) in the Main Text for a given choice of the parameters $k$, $D$, and $\beta$, with a matrix $\hat{A}$ determined by the arrangement of PAM sites in the given experiment. We maximize the joint log-likelihood 
\begin{equation}
\ln \mathcal{L}=\sum_{j=0}^5 \ln\mathcal{L}_j
\end{equation}
with respect to the three parameters and compute their uncertainties from the curvature of the log-likelihood. To facilitate a visual comparison, individual curves for each number of PAM sites and corresponding experimental data are shown in Fig.~\ref{fig:s1} (same as Fig.~2b in the Main Text, but with each experiment in a different panel).

\section{Sequence-dependent model}\label{sec:sequence}

In the model introduced in the Main Text, the binding energy of Cas9 to any triplet other than PAM is the same. In this Section, we introduce a model that relaxes this assumption and study its properties. To this aim, we define as "canonical PAM" a NGG triplet (where N stands for any base) and "non-canonical PAMs" the 15 possible triplets where either one or both G are replaced by other bases. This definition is motivated by the observation that the first "N" base of PAM does not to affect the binding energy of Cas9 \cite{bonomo2018physicist}. However, in principle, the binding energy of Cas9 with each non-canonical PAM can depend on the other two bases,  and the assumption made in the Main Text should be considered as a simplification.

We determine the binding energies of  non-canonical PAMs from  experimental results of the equilibrium occupancy of  off-target dsDNA bound by dCas9 \cite{boyle2017high}, a mutant of Cas9 that lacks the endonuclease capability. In the experiment, double strand DNA sequences containing the 20 bps main target and all possible replacement of the ``GG'' in the PAM are fixed in the flow cell. After 12 hours incubation using 10nM dCas9, the occupancy of the dsDNA sequences is measured (see Fig.~2S in \cite{boyle2017high}).

The measurement is performed after incubation, so that the system can be assumed to be at equilibrium. Every target DNA sequence can either be occupied by one Cas9 or empty. The occupancy $O$ is therefore expressed by the Fermi-Dirac distribution
\begin{equation}
  O_i=\frac{1}{1+e^{\epsilon_i-\mu}}
  \label{equation:FD}
\end{equation}
where $\epsilon_i$ is the binding energy of a particular triplet $i$ and $\mu$ is the chemical potential of dCas9. In the experiment, all canonical and non-canonical PAMs are followed by an identical 20bp target. Therefore, we expect differences in $\epsilon_i$ to depend on the different non-canonical PAMs only. The authors of Ref. \cite{boyle2017high} report the occupancy $O_i$ for all non-canonical PAMs relative to the canonical one. We call $\epsilon_T$ the binding energy of the specific target (i.e. the canonical PAM). We assume that $\epsilon_T$ is sufficiently negative so that the occupancy of the target is approximately equal to 1. Accordingly, we interpret the relative occupancies $O_i$ as absolute ones.

We use Eq.~\eqref{equation:FD} to eliminate $\mu$ and express the binding energy difference between a generic non-canonical PAM and the weakest non-canonical PAM (NTC) that we take as reference: 
\begin{equation}\label{eq:deltaeps}
 \Delta \epsilon_i= \epsilon_i-\epsilon_{NTC}=\ln\left[\frac{O_{NTC}(1-O_i)}{O_i(1-O_{NTC})}\right].
\end{equation}
Equation~\eqref{eq:deltaeps} permits to determine the binding energy difference from the experimental occupancy data. Results are presented in Table~\ref{energy table}. 

The diffusion and unbinding rates of the sequence-dependent model are defined from these energy differences as: 
 \begin{align}
     D_{n+1,n}&=D_{n-1,n}=D' e^{\Delta \epsilon_n}\nonumber\\ 
     k_n&=k' e^{\Delta \epsilon_n} , \label{eq:seqdepen}
 \end{align}
where we denoted the diffusion rate and the unbinding rate with $D'$ and $k'$, respectively, to distinguish them from the rate $D$ and $k$ appearing in the model presented in the Main Text. We run simulations of the sequence-dependent model using the binding energy differences in Table~\ref{energy table}. In this case, the free parameters are $D'$, $k'$, and the binding energy difference $\Delta \epsilon_T=\epsilon_T-\epsilon_{NTC}$ between the canonical PAM and the weakest non-canonical NTC. We fit these three parameters using the FRET data from \cite{globyte2019crispr}, following the same procedure described in Section I and using the specific DNA sequences that Ref.~\cite{globyte2019crispr} reports for each experiment. We obtain  $D'=160 s^{-1}, k'=6.57 s^{-1}$, and $\Delta \epsilon_T=-4.47$.

\begin{table}[htb!]
\center
\begin{tabular}{p{1cm}|c|c|c|c}
& G & A & C & T\\ \hline
G & & -2.61 & -1.12 & -1.42\\
\hline
A & -2.59 & -1.04 & -0.975 & -1.35\\
\hline
C & -1.22 & -1.40 & -1.35 & 0\\
\hline
T & -1.08 & -0.953 & -0.680 & -1.12\\
\end{tabular}
\label{energy table}
\caption{non-canonical PAM energies $\Delta \epsilon_i$. Rows represent the first nucleotide and columns for the nucleotide next to the ``N'' }
\end{table}

\begin{figure}[htb!]
\centering
\vspace*{0.1em}
\includegraphics[width=1.0\linewidth]{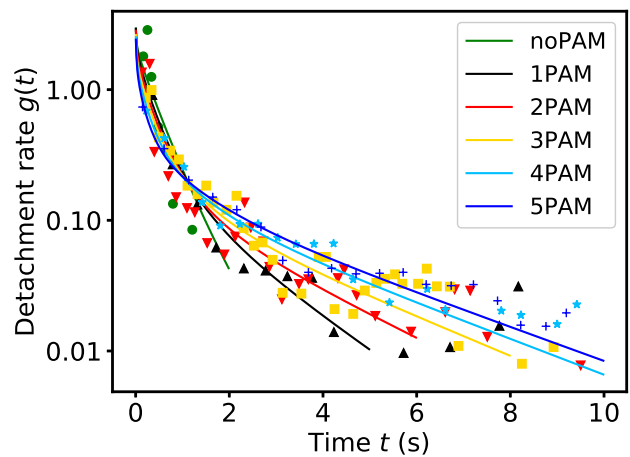}
\caption{Detachment rate $g(t)$ for $j=0\dots 5$ PAM sites predicted by the sequence-dependent model (lines) versus experimental measures from Ref. \cite{globyte2019crispr} (points). See Fig.~2b in the Main Text for comparison and more information. The fit returns a value of $\chi^2=280.4$,  compared with $\chi^2= 276.6$ in the model presented in the Main Text.}
\label{fig:s2}
\end{figure}

We now compare these parameters with those for the nearest-neighbor model presented in the Main Text. In the sequence-dependent model, the average energy of non-canonical PAM sites is $\epsilon_{av}=-1.26$ (see Table I). We now express the average diffusion rate between neighboring non-canonical PAM sites as

\begin{equation}
    \langle D_{n+1,n}\rangle = D' \langle e^{\Delta \epsilon_n}\rangle = 54 s^{-1}
\end{equation}
In contrast, in the model presented in the Main Text, we have $\langle D_{n+1,n}\rangle =D=52 s^{-1}$. The relative difference between these two values is about 4\%.

In the sequence-dependent model, we similarly have that the average unbinding rate from a non-canonical PAM is expressed by
\begin{equation}
    \langle k_n\rangle = k'  \langle e^{\Delta \epsilon_n}\rangle = 2.2 s^{-1}
\end{equation}
whereas in the model of the Main Text we have $\langle k_n\rangle=k=1.94 s^{-1}$. In this case, the relative discrepancy is 12.5\%.

Finally, in the sequence-dependent model, the energy difference between the canonical PAM and an average non-canonical PAM is equal to $\epsilon_T-\epsilon_{av}=-3.21$. This value is close to the estimated value $\beta=-3.34$ of the model in the Main Text, with a relative discrepancy of 4\%.

With these fitted parameters, we find that the sliding length in the one PAM case is equal to $\ell=5.2$ bp compared with  $6.2$ bp for the model in the Main Text.

\begin{figure}[htb!]
\centering
\vspace*{0.1em}
\includegraphics[width=1.0\linewidth]{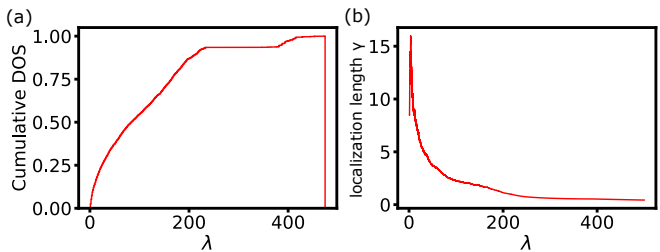}
\caption{(a) Cumulative density of states (DOS) and (b) localization length as function of $\lambda$ for the sequence-dependent model,  Eq.~\eqref{eq:seqdepen}, computed the transfer matrix method and Eqs.~(14) and~(15) in the Main Tex.  The DNA chain length is $N=5000$.}
\label{fig:5s}
\end{figure}

We also computed the localization length and the density of states for the sequence-dependent model in the disordered case, see Fig.~\ref{fig:5s}. For the sequence-dependent model, the maximum localization length is slightly larger than for the model in the Main Text ($\gamma\approx 15$ vs $\gamma\approx 10$, respectively). This difference should not be surprising, since the localization length is expected to be particularly sensitive to the distribution of the disorder. In any case, the qualitative result is confirmed, in the sense that both sliding lengths are much shorter than what it is observed in immunoprecipitation experiments.

We conclude from these comparisons that the physical picture resulting from the model presented in the Main Text is consistent with the one provided by this more detailed model.

\section{Regular versus disordered assortment of PAM Sites}\label{sec:regvsdisord}

\begin{figure}[htb!]
\centering
\vspace*{0.1em}
\includegraphics[width=0.9\linewidth]{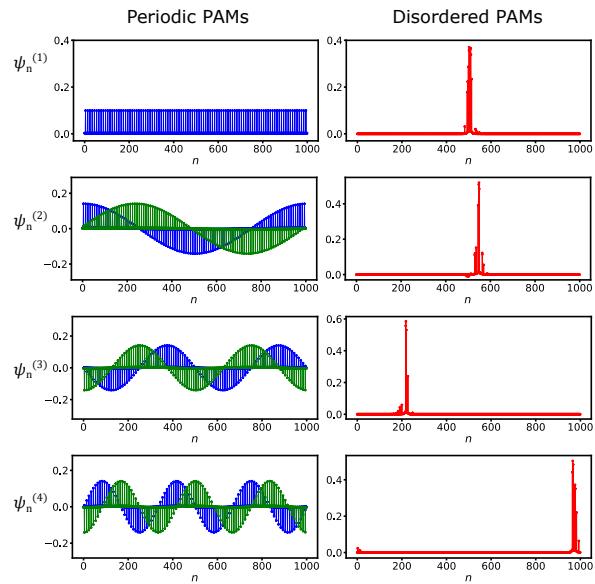}
\caption{Comparison of the first four eigenvectors of Cas9 sliding dynamics for (left) periodically spaced PAMs and (right) a disordered arrangement of PAM sites. In both cases, the length of the DNA chain is $N=1000$ and the average density of PAM sites is $1/10$. In the periodic case, the eigenvalues $\lambda_2$, $\lambda_3$, and $\lambda_4$ are associated with two degenerate eigenvectors (shown in blue and green in the figures). We obtained qualitatively similar results for closed boundary conditions (not shown). }
\label{fig:s6}
\end{figure}

Our interpretation of facilitated diffusion of Cas9 as a localization phenomenon leads to an interesting prediction. We expect eigenvectors characterizing Cas9 dynamics on a long DNA chain to be localized only if the PAM sites are arranged in a disordered fashion. If, instead, the PAM sites are regularly spaced, the eigenvectors should be extended as there is no disorder in this case. This prediction is confirmed in Fig.~\ref{fig:s6}. The figure shows that, in the case of regularly spaced PAM sites, the eigenvectors are characterized by peaks at each PAM site modulated by wave-like envelopes spanning the entire system size.

\section{Hopping model}\label{sec:hopping}

\begin{figure}[htb!]
\centering
\vspace*{0.1em}
\includegraphics[width=0.9\linewidth]{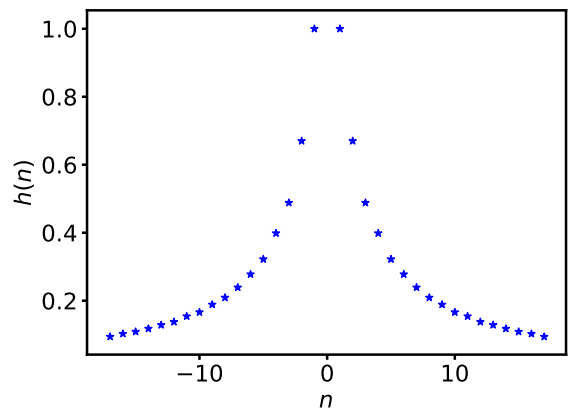}
\caption{Plot of the hopping distribution $h(n)$ versus $n$ for $\alpha=1$. The distribution $h(n)$ is normalized so that $h(1)=1$ and truncated at $n=17$ for computational convenience.}
\label{fig:s3}
\end{figure}

The hopping distribution $h(n)$ can be estimated from the solution of a diffusion equation in cylindrical coordinates \cite{lomholt2009facilitated}. The assumption of cylindrical symmetry is justified as far as we limit ourselves to hopping at distances much shorter than the DNA persistence length, which is on the order of 150 base pairs. On these short distances, the DNA double helix can be regarded as a straight cylinder.

\begin{figure}[htb!]
\centering
\vspace*{0.1em}
\includegraphics[width=0.9\linewidth]{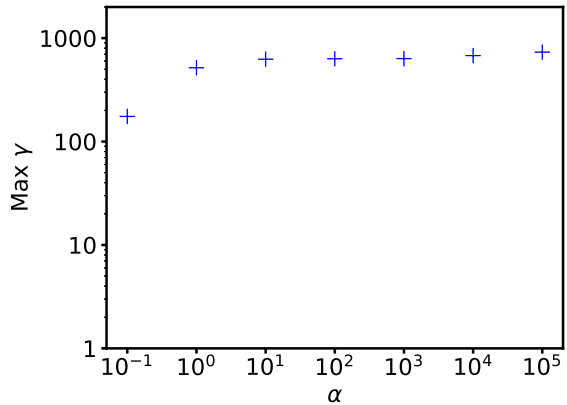}
\caption{Maximum localization length in the spectrum as a function of $\alpha$. For each value of $\alpha$, the maximum localization length is computed by direct diagonalization (as in Fig.~4d of the Main Text).}
\label{fig:s4}
\end{figure}

We consider the probability $W(n,t)$ of a protein to rebind at coordinate $n$ at time $t$, given that it detached at position $n=0$ and $t=0$. From the diffusion equation in cylindrical coordinates, the authors of Ref. \cite{lomholt2009facilitated} obtains the Fourier-Laplace transform 
\begin{equation}
\tilde{W}(q,u)=\int_0^\infty dt  ~e^{-ut} \int_{-\infty}^\infty dn~e^{iqn} W(n,t)  .
\end{equation}
In particular, the Fourier-Laplace transform calculated in $u=0$ yields the Fourier transform of the integrated probability of hopping to a given distance at any time: 
\begin{align}
\tilde{W}(q,0)&=\int_0^\infty dt \int_{-\infty}^\infty dn ~ e^{iqn} W(n,t) \nonumber\\
&=\left[1+\frac{2\pi \alpha|q|rK_1(|q|r)}{(K_0(|q|r))}\right]^{-1} ,
  \label{equation:W(q)}
\end{align} 
where $r=3$ is the DNA radius, measured in unit of the base pair distance, $K_j(n)$ is the modified Bessel function of the second kind, and $\alpha=1$ is the ratio between the 3D diffusion coefficient and the non-specific binding rate. For $n>1$, we compute $h(n)$ by a numerical inverse Fourier transform of $\tilde{W}(q,0)$, truncated at a distance $n_{\max}=17$. This maximum distance is chosen for computational convenient and is consistent with the assumption of cylindrical symmetry, as previously discussed. The hopping distribution $h(n)$ for $\alpha=1$, normalized such that $h(1)=1$, is shown in Fig.~\ref{fig:s3}. The localization length for the hopping model for $\alpha=1$ is shown in Fig.~4 of the Main Text. We found qualitatively similar results for the sliding length for $\alpha$  ranging from $0.1$ to $10^{5}$, see Fig.~\ref{fig:s4}.

\bibliography{localization}

\end{document}